\documentclass[12pt]{article}
\usepackage{rotating}
\usepackage{epsfig,amssymb}
\usepackage{ulem}
\usepackage{amsmath}
\input{epsf}

\def\laq{\raise 0.4 ex \hbox{$<$}\kern -0.8 em\lower 0.62 ex\hbox{$\sim$}}
\def\gaq{\raise 0.4 ex \hbox{$>$}\kern -0.7 em\lower 0.62 ex\hbox{$\sim$}}

\def\beq{\begin{equation}}
\def\eeq{\end{equation}}
\def\beqa{\begin{eqnarray}}
\def\eeqa{\end{eqnarray}}

\def\to{\rightarrow}

 \def\frac#1#2{{\textstyle{{#1}\over {#2}}}}
 \def\lsim{\mathrel{\rlap{\lower4pt\hbox{\hskip1pt$\sim$}}
    \raise1pt\hbox{$<$}}} \def\gsim{\mathrel{\rlap{\lower4pt\hbox{\hskip1pt$\sim$}}
    \raise1pt\hbox{$>$}}}
\def\sqr#1#2{{\vcenter{\vbox{\hrule height.#2pt
         \hbox{\vrule width.#2pt height#1pt \kern#1pt
         \vrule width.#2pt}
         \hrule height.#2pt}}}}

 \def\frac#1#2{{\textstyle{{#1}\over
{#2}}}} 
\def\lsim{\mathrel{\rlap{\lower4pt\hbox{\hskip1pt$\sim$}}
\raise1pt\hbox{$<$}}}
\def\gsim{\mathrel{\rlap{\lower4pt\hbox{\hskip1pt$\sim$}}
\raise1pt\hbox{$>$}}} \def\sqr#1#2{{\vcenter{\vbox{\hrule height.#2pt
\hbox{\vrule width.#2pt height#1pt \kern#1pt \vrule width.#2pt} \hrule
height.#2pt}}}}

\def\beq{\begin{equation}} \def\eeq{\end{equation}}
\def\beqa{\begin{eqnarray}} \def\eeqa{\end{eqnarray}}

\def\gappeq{\mathrel{\rlap {\raise.5ex\hbox{$>$}} {\lower.5ex\hbox{$\sim$}}}}

\def\lappeq{\mathrel{\rlap{\raise.5ex\hbox{$<$}}
{\lower.5ex\hbox{$\sim$}}}}

\usepackage{color}

\begin{document}
\pagestyle{plain}

\begin{flushright}
%DF/IST--2.2007\\
February 2024
\end{flushright}
\vspace{15mm}

\begin{center}

{\Large\bf Time Crystals} 

{\Large\bf and} 

{\Large\bf Phase-Space Noncommutative Quantum Mechanics}

\vspace*{1.0cm}

Orfeu Bertolami$^{1,2}$\\
\vspace*{0.2cm}
{$^{1}$ Departamento de F\'{\i}sica e Astronomia, Faculdade de Ci\^encias,
Universidade do Porto, \\
Rua do Campo Alegre s/n, 4169-007 Porto, Portugal}\\

{$^{2}$ Centro de F\'{\i}sica das Universidades do  Minho e do Porto,
Rua do Campo Alegre s/n, 4169-007 Porto, Portugal}\\

\vspace*{0.2cm}
and

A. E. Bernardini\

Departamento de F\'{\i}sica, Universidade Federal de S\~ao Carlos, \\ PO Box 676, 13565-905, S\~ao Carlos, SP, Brasil.

%{

\vspace*{2.0cm}
\end{center}

\begin{abstract}
\noindent
We argue that time crystal properties naturally arise from phase-space 
noncommutative quantum mechanics. In order to exemplify our point we consider 
the 2-dimensional noncommutative quantum 
harmonic oscillator and show that it exibihits
periodic oscillations that can be identified as time crystals. 

\end{abstract}

\vfill
\noindent\underline{\hskip 140pt}\\[4pt]
\noindent
{E-mail addresses: orfeu.bertolami@fc.up.pt; alexeb@ufscar.br}

\noindent
{Based on talk presented by one of us (O.B.) at the Third Minkowsky Meeting on the Foundations of the Spacetime Physics at Albena, Bulgaria, 11-14 September 2023.} 
\newpage

\section{Introduction}
\label{sec:introduction}

In this contribution we review the arguments presented in Ref. \cite{BB22} where it was 
shown that time crystal features arise in the context of the phase-space noncommutative 2-dimensional quantum harmonic oscillator. As discussed in the following this is yet another new property emerging from phase-space noncommutative quantum mechanics (PSNCQM).

Time crystals are time-periodic self-organized structures that presumably arise due to the spontaneous breaking of time translation symmetry \cite{W1,W2}. They are analogous to spatial crystal lattices that form when the spontaneous breaking of space translation symmetry takes place \cite{PR2018}. 
Time crystal features were claimed to appear in ultra-cold atoms \cite{Sacha15,Smits18} and spin-based solid state systems \cite{Khemani16,Else16,Pal,Rovny,Kyprianidis,Randall}, through which it has been argued that periodically driven systems exhibit a discrete time symmetry \cite{Zhang,Choi}.
In fact, these experiments suggest that novel phases of matter do exist \cite{Rovny,Kyprianidis,Randall} which exhibit a discrete time translation symmetry hinting the breakdown of the continuous time translation symmetry, $\hat{\mathcal{T}}_H\equiv e^{-i\hat{H}t}$. For a contextualization of time crystals with respect to the research on the physics of time, see, for instance, Ref. \cite{OB22}.
 
As is well known, if a time-independent system driven by a time-independent Hamiltonian, $H$, is prepared in an eigenstate $\vert\psi_n\rangle$, such that $H \vert\psi_n\rangle = E_n \vert\psi_n\rangle$, for the energy eigenvalue $E_n$, in the context of quantum mechanics (QM) the probability density at a fixed position in the configuration space is also time-independent. Nevertheless, the mentioned experiments suggest that time crystals exist and thus, $[\hat{H},\,\rho_n]\equiv[\hat{\mathcal{T}}_H,\,\rho_n]\neq 0$ for $\rho_n = \vert\psi_n\rangle\langle\psi_n\vert$.

As originally argued \cite{W1}, this would correspond to a spontaneous breakdown of time translation symmetry followed by a non-stationary behaviour of the eigensystem solutions.
However, a no-go theorem \cite{Watanabe}, based on the time-dependent correlation functions of the order parameter, rules out the possibility of time crystals defined in this way for the ground state and  for a canonical ensemble of a general Hamiltonian. We argue that the emergence of a non-stationary behaviour, and its connection with time crystal properties can be explained in terms of both position and/or momentum noncommutativity in the phase-space \cite{BB22}, in opposition to the {\it ab initio} breaking symmetry assumptions proposed in Refs.  \cite{W1,W2}.

\section{Phase-Space Noncommutative Quantum Mechnics}
\label{sec:PSNCQM}

We present now some of the main features of PSNCQM. Noncommutativity was firstly considered in the space coordinate domain as a way to regularize quantum field theories \cite{Snyder47} and subsequently in string theory \cite{Connes,Douglas,Seiberg,Nekrasov01}. The PSNCQM extension \cite{Gamboa,06A,Rosenbaum,Jing,Bastos,08A,09A}, considered here, can be formulated in terms of the Weyl-Wigner-Groenewold-Moyal (WWGM) framework \cite{Groenewold,Moyal,Wigner}, supported by a $2n$-dimensional phase-space deformed Heisenberg-Weyl algebra, where position and momentum operators,  $\hat{q}_i$ and $\hat{p}_j$, obey the commutation relations,
\begin{equation}
[ \hat{q}_i, \hat{q}_j ] = i \theta_{ij} , \hspace{0.5 cm} [ \hat{q}_i, \hat{p}_j ] = i \hbar \delta_{ij} ,
\hspace{0.5 cm} [ \hat{p}_i, \hat{p}_j ] = i \eta_{ij},\label{EQEq31}
\end{equation}
where $ i,j= 1, ... ,d$, and $\eta_{ij}$ and $\theta_{ij}$ are the entries of invertible antisymmetric real constant ($d \times d$) matrices, $ {\bf \Theta}$ and ${\bf N}$, such that an equally invertible matrix, ${\bf \Sigma}$, with $\Sigma_{ij} \equiv \delta_{ij} + \hbar^{-2} \theta_{ik} \eta_{kj}$, exists, which demands that $\theta_{ik}\eta_{kj} \neq -\hbar^2 \delta_{ij}$.
Of course, given that $\eta_{ij} \neq 0,\,\theta_{ij} \neq 0$, the relations from Eq.~\eqref{EQEq31} can affect the symmetries related to conserved quantities associated to quantum operators, $\hat{\mathcal{O}}$, for which
$d \langle\hat{\mathcal{O}}\rangle/dt = i\hbar^{-1}\,\langle[\hat{H},\,\hat{\mathcal{O}}]\rangle =0$.

In this context, the key issue is if quantum operators identified by $\hat{\mathcal{O}} \to \hat{\mathcal{O}}(\{\hat{q}_i,\hat{p}_i\})$ do present a time crystal behaviour arising from the breakdown of time translational symmetries, $\langle[\hat{H},\,\hat{\mathcal{O}}(\{\hat{q}_i,\hat{p}_i\})]\rangle \neq 0$, in opposition to usual QM.
In order to investigate this point, the NC algebra, Eq.~\eqref{EQEq31} can be mapped into the Heisenberg-Weyl algebra through the linear Seiberg-Witten (SW) transformation \cite{Seiberg}, 
\begin{equation}
 \hat{q}_i = A_{ij} \mathit{\hat{Q}}_j + B_{ij} \hat{\Pi}_j, \hspace{1 cm}
 \hat{p}_i = C_{ij} \mathit{\hat{Q}}_j + D_{ij} \hat{\Pi}_j,
\label{EQEq35}
\end{equation}
where $A_{ij}, B_{ij}, C_{ij}$ and $D_{ij}$ are real entries of constant matrices, ${\bf A}, {\bf B}, {\bf C}$ and ${\bf D}$. 
In this case, one recovers the algebra of ordinary QM,
\begin{equation}
[ \hat{Q}_i, \hat{Q}_j ] = 0, \quad [ \hat{Q}_i, \hat{P}_j ]
= i \hbar \delta_{ij} , \hspace{0.5 cm} [ \hat{P}_i, \hat{P}_j ] = 0,
\label{EQNCeq}
\end{equation}
through the following matrix equation constraints \cite{Rosenbaum},
${\bf A} {\bf D}^T - {\bf B} {\bf C}^T = {\bf I}_{d \times d}$, ${\bf A} {\bf B}^T - {\bf B} {\bf A}^T = \hbar^{-1} {\bf \Theta}$, and ${\bf C} {\bf D}^T - {\bf D} {\bf C}^T = {\hbar^{-1}} {\bf N}$,
where the superscript $T$ denotes matrix transposition.

From the WWGM framework \cite{Rosenbaum,Bastos} for the algebra, Eq.~\eqref{EQEq31}, it is possible to show 
that the resulting quantum mechanical extensions have some striking features which include putative violations of the Robertson-Schr\"odinger uncertainty relation \cite{Bastos001,Bastos002}, quantum correlations and information collapse in gaussian quantum systems \cite{Bernardini13B,Bernardini13B2,RSUP1,RSUP2,RSUP3}, new regularizing features in minisuperspace quantum cosmology models \cite{Bastos001,Bastos003} and in black-hole physics \cite{Bastos03B,Bastos004,Bastos005}, putative violations of the Equivalence Principle \cite{Bastos006,Leal01} and, likewise ordinary QM, non-locality properties that can be captured by the Bell operator \cite{Bastos007}.
In fact, the generalized WWGM star-product, the extended Moyal bracket and the noncommutative (NC) Wigner function framework ensure that observables are independent of any particular choice of the SW map \cite{Bastos}.

\section{The Noncommutative 2-dimensional Quantum Harmonic Oscillator and the Emergence of Time Crystal Behaviour}
\label{sec:2dHO}

Aiming to exemplify the emerging time crystal behaviour we consider the $2$-dimensional harmonic oscillator in the PSNCQM  \cite{Bernardini13A} with Hamiltonian,
\begin{equation}\label{EQNCCC}
\hat{H}_{HO}(\hat{\mathbf{q}},\hat{\mathbf{p}}) = {\hat{\mathbf{p}}^2 \over 2m} + \frac{1}{2}m \omega^2 \hat{\mathbf{q}}^2,
\end{equation}
on the NC ``$x-y$'' plane, with position and momentum satisfying the NC algebra, Eq.~\eqref{EQEq31}, now with $i,j=1,2$, $\theta_{ij} = \theta \epsilon_{ij}$ and $\eta_{ij} = \eta \epsilon_{ij}$, where $\epsilon_{ij}$ is the $2$-dimensional Levi-Civita tensor. The map to commutative operators is given by
\begin{equation}
\mathit{\hat{Q}}_i = \mu \left(1 - {\theta \eta\over\hbar^2} \right)^{- 1 / 2} \left( \hat{q}_i + {\theta\over 2 \lambda \mu \hbar} \epsilon_{ij} \hat{p}_j \right),\quad
\hat{\Pi}_i = \lambda \left(1 - {\theta \eta\over\hbar^2} \right)^{-1 / 2} \left( \hat{p}_i-{\eta\over 2 \lambda \mu \hbar} \epsilon_{ij} \hat{q}_j \right),
\label{EQSWinverse}
\end{equation}
in terms of the SW map,
\begin{equation}
 \hat{q}_i = \lambda \mathit{\hat{Q}}_i - {\theta\over2 \lambda \hbar} \epsilon_{ij} \hat{\Pi}_j, \hspace{0.5 cm}\hspace{0.5 cm} \hat{p}_i = \mu \hat{\Pi}_i + {\eta\over 2 \mu \hbar} \epsilon_{ij} \mathit{\hat{Q}}_j~,
\label{EQSWmap}
\end{equation}
which is invertible for $\theta\eta \neq \hbar^2$, and the parameters $\lambda$ and $\mu$ satisfying the condition
\begin{equation}
{\theta \eta \over 4 \hbar^2} = \lambda \mu ( 1 - \lambda \mu ).
\label{EQconstraint}
\end{equation}

The Hamiltonian in terms of the commutative variables, $\mathit{\hat{Q}}_i$ and $\hat{\Pi}_i$, reads \cite{Bernardini13A}
\begin{equation}
\hat{H}_{HO}(\hat{\mbox{\bf \em Q}},\hat{\mathbf{\Pi}}) = \alpha^2\hat{\mbox{\bf \em Q}}^2 +\beta^2\hat{\mathbf{\Pi}}^2 + \gamma \sum_{i,j = 1}^2{\epsilon_{ij}\hat{\Pi}_i \mathit{\hat{Q}}_j},
\label{EQHamilton}
\end{equation}
where  ${\alpha}^2 \equiv m  \omega^2\lambda^2/2 + \eta^2/(8m  \hbar^2\mu^2)$, ${\beta}^2 \equiv {\mu^2/(2m)} + {m \omega^2 \theta^2/(8 \hbar^2 \lambda^2)}$, and ${\gamma} \equiv m \omega^2{\theta}/({2\hbar}) + {\eta}/({2m\hbar})$, from which one obtains the following set of coupled equations of motion,
\begin{eqnarray}
\dot{\Pi}_i &=& -\frac{i}{\hbar} \langle\left[\hat{\Pi}_i,\,\hat{H}_{HO}\right]\rangle = -2 \alpha^2\,\mathit{Q}_i - \gamma\,\varepsilon_{ji}\Pi_j,\nonumber\\
\dot{\mathit{Q}}_i &=& -\frac{i}{\hbar} \langle\left[\hat{\mathit{Q}}_i,\,\hat{H}_{HO}\right]\rangle = ~~2 \beta^2\,\Pi_i - \gamma\,\varepsilon_{ji}\mathit{Q}_j,
\label{EQeqs01}
\end{eqnarray}
with $\mathit{Q}_i \equiv \langle \hat{\mathit{Q}}_i \rangle$ and ${\Pi}_i \equiv \langle \hat{\Pi}_i \rangle$. In this case, $\mbox{\bf \em Q} = (\mathit{Q}_1,\,\mathit{Q}_2)$ and $\mathbf{\Pi}= ({\Pi}_1,\,{\Pi}_2)$ may be interpreted as the dynamical variables within the WWGM formalism for which the solutions are given by \cite{Bernardini13A}
\small
\begin{eqnarray}
\mathit{Q}_1(t)&=& x\,\cos(\Omega t)\cos(\gamma t) + y\,\cos(\Omega t)\sin(\gamma t)
 +\frac{\beta}{\alpha}\left[\pi_y\,\sin(\Omega t)\sin(\gamma t) + \pi_x\,\sin(\Omega t)\cos(\gamma t)
\right],\nonumber\\
\mathit{Q}_2(t)&=& y\,\cos(\Omega t)\cos(\gamma t) - x\,\cos(\Omega t)\sin(\gamma t)
 -\frac{\beta}{\alpha}\left[\pi_x\,\sin(\Omega t)\sin(\gamma t) - \pi_y\,\sin(\Omega t)\cos(\gamma t)
\right],\nonumber\\
\Pi_1(t)&=& \pi_x\,\cos(\Omega t)\cos(\gamma t) + \pi_y\,\cos(\Omega t)\sin(\gamma t)
 -\frac{\alpha}{\beta}\left[y\,\sin(\Omega t)\sin(\gamma t) + x\,\sin(\Omega t)\cos(\gamma t)
\right],~~\nonumber\\
\Pi_2(t)&=& \pi_y\,\cos(\Omega t)\cos(\gamma t) - \pi_x\,\cos(\Omega t)\sin(\gamma t)
 +\frac{\alpha}{\beta}\left[x\,\sin(\Omega t)\sin(\gamma t) - y\,
\sin(\Omega t)\cos(\gamma t)
\right],~~
\label{EQsolutions}
\end{eqnarray}
\normalsize
where $x,\, y,\,\pi_x,$ and $\pi_y$ are arbitrary parameters, and 
\begin{equation}
\Omega = 2 \alpha \beta = \sqrt{(2\lambda\mu - 1)^2\omega^2 + \gamma^2} = \sqrt{\omega^2 +\gamma^2 - {\theta \eta \over \hbar^2}},
\label{EQeq37}
\end{equation}
with $\lambda$ and $\mu$ being eliminated by the constraint Eq.~\eqref{EQconstraint}.
Of course, if one sets $\theta=\eta = 0$, and therefore $\gamma = 0$, one recovers the solutions for the $2$-dimensional harmonic oscillator with uncoupled $x-y$ coordinates and $\Omega = \omega$.
For $\theta,\, \eta \neq 0$, the above results lead to two decoupled time-invariant quantities,
\begin{eqnarray}
\sum_{i=1}^2{\left(\frac{\alpha}{\beta} \mathit{Q}_i(t)^2 + \frac{\beta}{\alpha}\Pi_i(t)^2\right)} &=& \frac{\alpha}{\beta} (x^2 + y^2) + \frac{\beta}{\alpha}(\pi_x^2 + \pi_y^2),\nonumber\\
\sum_{i,j=1}^2{\left(\epsilon_{ij} \mathit{Q}_i(t)\,\Pi_j(t)\right)} &=& x\,\pi_y - y \,\pi_x.
\label{EQinvariant}
\end{eqnarray}
The changes introduced by the NC variables can be evinced by setting $\pi_x = \pi_y = \sqrt{\alpha\hbar/2\beta}$, and $x = y = \sqrt{\beta\hbar/2\alpha}$, so that the associated $x$ and $y$ translational energy contributions evolve as 
\begin{equation}\small
E_i =\alpha\beta {\left(\frac{\alpha}{\beta} \mathit{Q}_i(t)^2 + \frac{\beta}{\alpha}\Pi_i(t)^2\right)} = {\hbar\Omega \over 2} \left(1 - (-1)^i\,\sin(2\gamma t)\right),
\label{EQenergy}
\end{equation}\normalsize
with $i = 1,\,2$, from which a typical low frequency $\gamma$-dependent beating behaviour is encountered \cite{Bernardini13A}.
Such a time-dependent periodic modification is a new feature of the NC harmonic oscillator ground state. The {\it stargen}functions for the Hamiltonian, Eq.~(\ref{EQHamilton}), are obtained from the {\it stargen} value equation,
\begin{equation}
H^W_{HO} \star \rho_{n_{\tiny 1},n_{\tiny 2}}^W (\mbox{\bf \em Q},\mathbf{\Pi}) = 
E_{n_{\tiny{1}},n_{\tiny{2}}}\,\rho^W_{n_{\tiny{1}},n_{\tiny{2}}} (\mbox{\bf \em Q},\mathbf{\Pi}),
\label{EQhelp01}
\end{equation}
where $W (\mbox{\bf \em Q},\mathbf{\Pi})$ is the eigenstate associated Wigner function, from which one has \cite{Rosenbaum},
\begin{equation}
\rho_{n_{\tiny{1}},n_{\tiny{2}}}^W (\mbox{\bf \em Q},\mathbf{\Pi}) = {(-1)^{n_1+n_2} \over \pi^2\hbar^{2}}\exp\left[{-\frac{1}{\hbar}\left(\frac{\alpha}{\beta}\mbox{\bf \em Q}^2 + \frac{\beta}{\alpha}{\mathbf{\Pi}}^2\right)}\right] \, L^0_{n_1} \left(\Omega_{+}/\hbar\right) \,L^0_{n_2}\left(\Omega_{-}/\hbar\right),
\label{EQLague01}
\end{equation}
where $L^0_n$ are the associated Laguerre polynomials, $n_1$ and $n_2$ are non-negative integers, and
\begin{equation}
{\Omega}_{\pm} = {\alpha\over\beta}\mbox{\bf \em Q}^2 + {\beta\over\alpha}\mathbf{\Pi}^2 \mp 2 \sum_{i,j = 1}^2{\left(\epsilon_{ij}\Pi_i \mathit{Q}_j\right)},
\end{equation}
such that the energy spectrum is given by
$E_{n_{\tiny 1},n_{\tiny 2}} = \hbar\left[2\alpha\beta(n_1 + n_2 + 1) + \gamma (n_1 - n_2)\right]$.

\begin{figure}[h!]
\includegraphics[width= 9 cm]{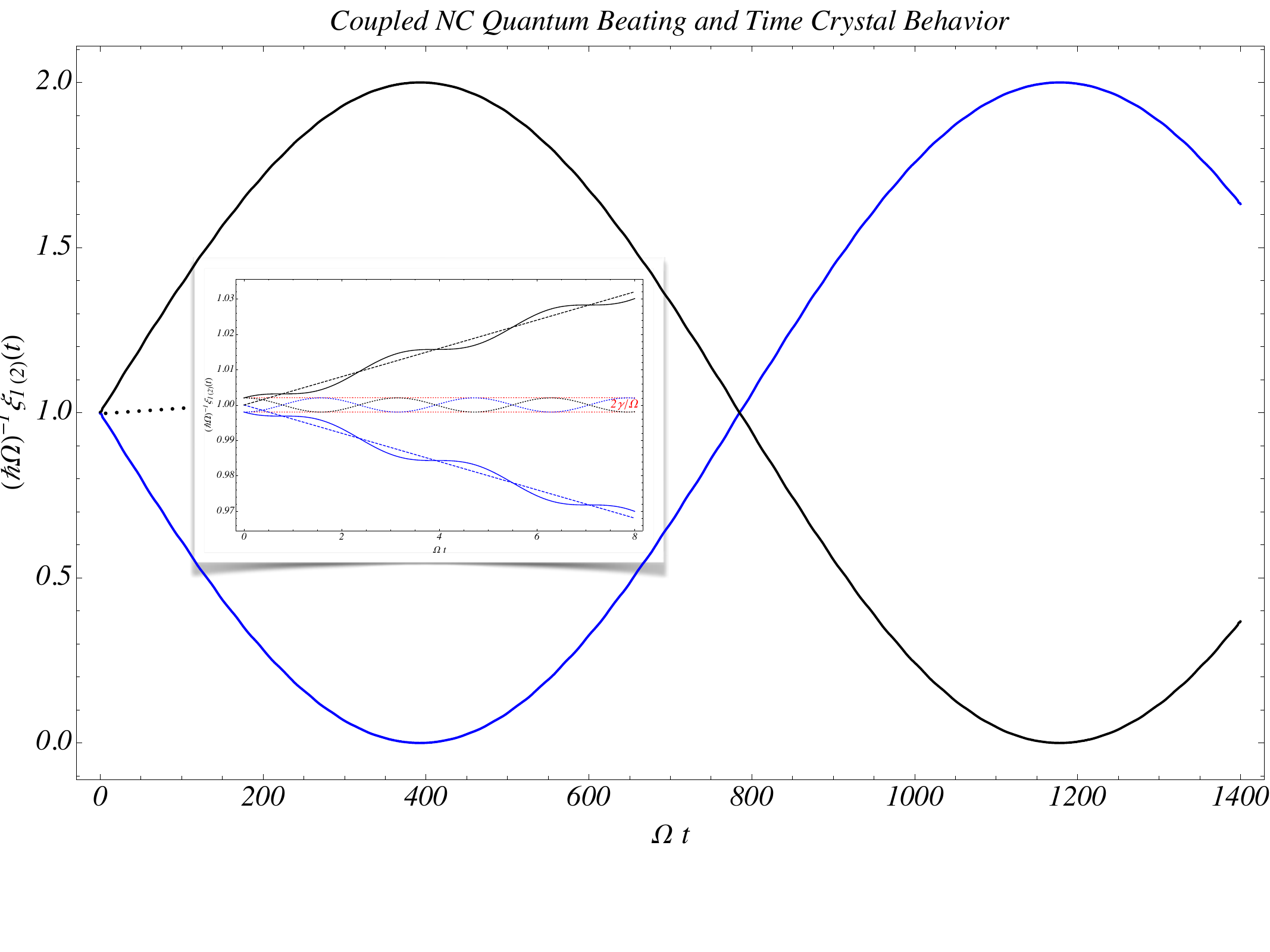}
\vspace{-1. cm}\renewcommand{\baselinestretch}{1}
\caption{(Colour on line) Dimensionless NC associated energies , $(\hbar\Omega)^{-1}{\xi}_{1(2)}$ (black (blue) line) as function of $\Omega t$, for $\gamma /\Omega=0.002$. Decoupled $\gamma$-frequency NC quantum beating (dashed lines) and $\Omega$-frequency time crystal behaviour (dotted lines) for $\gamma t \gtrsim 0$ are identified in the zoom in window. Figure from Ref. \cite{BB22} }
\label{EQTC0004}
\end{figure}

\begin{figure}[h!!]
\includegraphics[width= 9 cm]{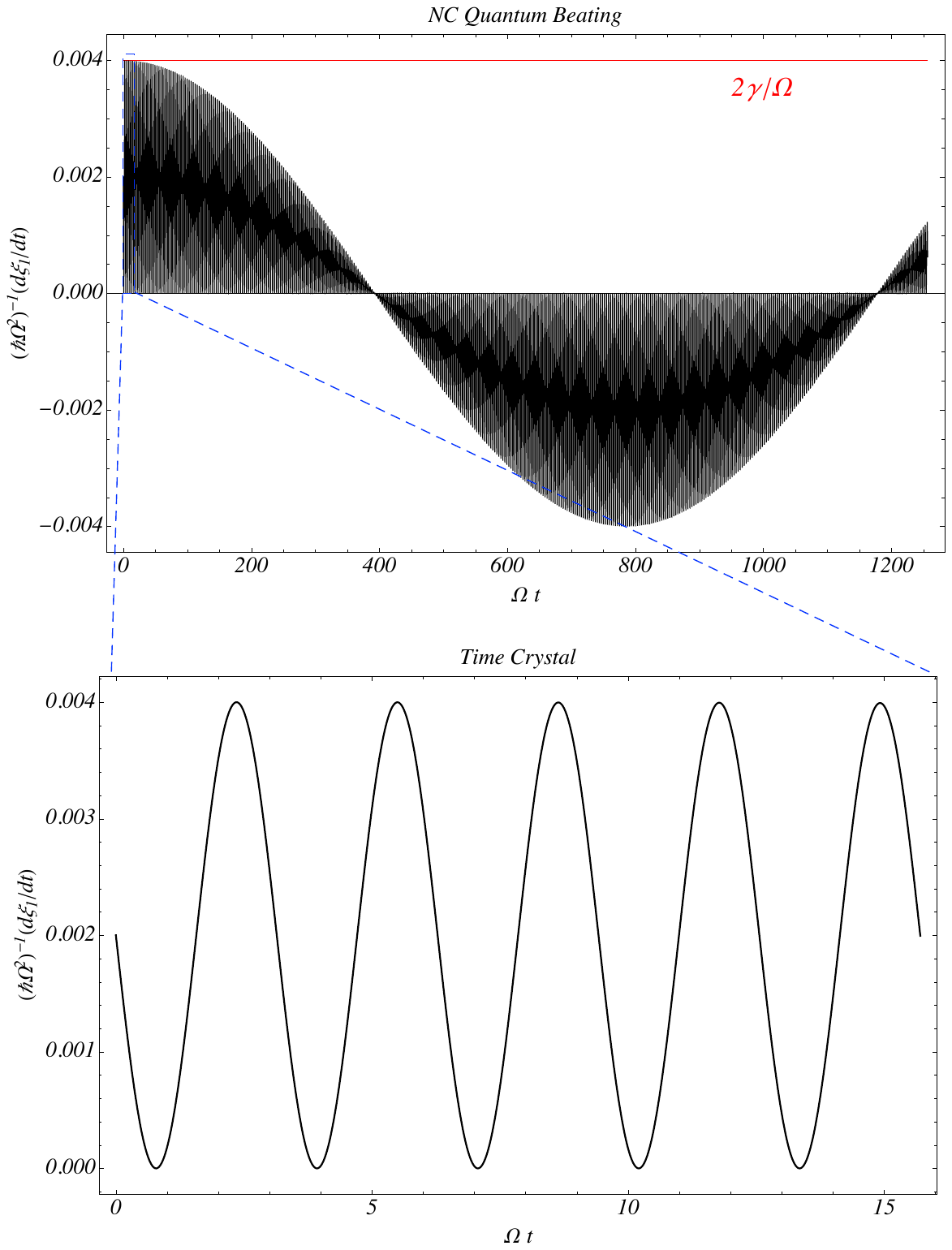}
\vspace{-1. cm}\renewcommand{\baselinestretch}{1}\caption{(Colour on line) Dimensionless time derivative, $(\hbar\Omega^2)^{-1}\dot{\xi}_1$ as function of $\Omega t$, for $\gamma /\Omega=0.002$. The NC beating behaviour is depicted by the first plot and the time crystal periodic behaviour, driven by the amplitude modulation, $\hbar \gamma \Omega(\equiv \gamma/\Omega \times \hbar \Omega^2)$ (red line) is depicted in zoom in plot. Figure from Ref. \cite{BB22}}
\label{EQTC0001}
\end{figure}

It has been shown that the $2$-dimensional harmonic oscillator on the NC plane approaches the classical limit, and exhibits well-marked quantum effects such as state swapping, quantum beating, and some extent of loss of quantum coherence. These properties are not due to extrinsic or artificial time-dependent effects, but due to the entanglement \cite{Bernardini13B} induced by NC ``$x$'' and ``$y$'' Hilbert spaces mapped by the time-independent Hamiltonian. From Eq.~\eqref{EQHamilton} one sees that quantum states associated to $x$ and $y$ degrees of freedom are no longer independent. This differs from the standard quantum mechanical configuration, for which $x$ and $y$ modes are each of them associated to decoupled stationary behaviour \cite{Bernardini13A}.
In order to clarify the relation with the time crystal behaviour, one should get back to the Hamiltonian Eq.~\eqref{EQNCCC} and examine the contributions of $\{\hat{q}_1,\,\hat{p}_1\}$ and $\{\hat{q}_2,\,\hat{p}_2\}$ to the energy and to the eigenstates. Indeed, identifying the associated energy of each $i$-sector ($i=1,\,2$)  as
\begin{equation}\label{EQNCCC2}
\hat{\xi}_i = {\hat{p}_i^2 \over 2m} + \frac{1}{2}m \omega^2 \hat{q}_i^2,
\end{equation}
from standard QM, one would have $\dot{\xi}_i = i\hbar^{-1} \langle [H,\, \hat{\xi}_i]\rangle = 0$ (with $\langle \hat{\xi}_i\rangle \equiv {\xi}_i$).
However, after recasting $\{\hat{q}_i,\,\hat{p}_i\}$ in terms of the SW map, Eqs.~\eqref{EQSWinverse}-\eqref{EQSWmap}, with $\Omega$, $\omega$ and $\gamma$ constrained by Eq.~\eqref{EQeq37}, one finds an unexpected non-stationary behaviour for each of the energy contributions,
\begin{eqnarray}\label{EQNCCC3}
\xi_i(t) &=& 
{\hbar \Omega \over 2}
\left\{1 - (-1)^i
\left[
\sqrt{1-\frac{\omega ^2}{\Omega ^2}}\left(\cos (2 \gamma t)\, \cos (2 \Omega t )-
\frac{\gamma}{\Omega} \sin (2 \gamma t)\, \sin (2 \Omega t )\right)\right.\right.\\
&&\qquad\qquad\qquad\qquad\qquad\qquad\qquad\qquad\qquad\qquad\qquad\left.\left.+\frac{\omega}{\Omega}\sqrt{1-\frac{\gamma ^2}{\Omega ^2}}\sin (2 \gamma t)
\right]\right\},\nonumber
\end{eqnarray}
from which arise the time crystal non-stationary behaviour driven by $\Omega$ and a beating behaviour driven by $\gamma$. This is depicted in Fig.~\ref{EQTC0004}; if either $\theta$ or $\eta$ vanishes, one has $\Omega^2 = \omega^2 +\gamma^2$ and
\small\begin{equation}
\xi_i(t)=
{\hbar \Omega \over 2}
\left\{1 - (-1)^i
\left[
\frac{\gamma}{\Omega}\left(\cos (2 \gamma t)\, \cos (2 \Omega t )-
\frac{\gamma}{\Omega} \sin (2 \gamma t)\, \sin (2 \Omega t )\right)+\left(1-\frac{\gamma ^2}{\Omega ^2}\right)\sin (2 \gamma t)
\right]\right\}.
\end{equation}\normalsize
For the arbitrary choice of $\gamma /\Omega=0.002$, it is shown in the smaller window of Fig.~\ref{EQTC0004}, the energy decoupled $\gamma$-frequency NC quantum beating (dashed lines) and the externally driven $\Omega$-frequency time crystal behaviour (dotted lines) for $\gamma t \gtrsim 0$.
In Fig.~\ref{EQTC0001} the time derivative of the energy is depicted, from which the magnitude of the time crystal oscillating behaviour can be quantified. From Eq.~\eqref{EQNCCC3}, the externally driven oscillation amplitude, $\hbar \Omega/2$, is modulated by a factor $ \gamma/\Omega$.

Given that the corrections due to the $\gamma$ parameter are small, the beating oscillations are presumably difficult to measure. On the other hand, an effect is acessible for $\gamma \ll \Omega$ and $\gamma t \gtrsim 0$ implying that 
\begin{equation}\label{EQNCCC4}
\xi_i(t)\approx 
{\hbar \Omega \over 2}\left[1 - (-1)^i \frac{\gamma}{\Omega} \left( 2\, \Omega t+\cos (2 \Omega t )\right)\right],
\end{equation}
at first order in $\gamma$. In this case,
\begin{equation}\label{EQNCCC5}
\dot{\xi}_i(t)\approx 
(-1)^{i+1}\hbar \gamma\Omega\left[1- \sin (2 \Omega t)\right] ,
\end{equation}
a time crystal periodic behaviour arise with a measurable energy time derivative oscillation amplitude, $\hbar \gamma \Omega$, driven by both the NC parameter, $\gamma$, and the external oscillation frequency $\Omega \sim \omega$.

It should be added that the states of the $2$-dimensional NC quantum harmonic oscillator here examined satisfy the no-cloning and no-deleting theorems without additional constraints \cite{Leal02}, given that these theorems depend only on the unitarity of QM, which is shared by PSNCQM. From these results it follows that some specific features of the system studied here cannot be ruled out on the basis of the same no-go theorems that render the Wilczek's hypothesis untenable \cite{Watanabe}.

Let us close this section pointing out that, as discussed in Ref. \cite{BB22} ,the effects presented above can also be inferred from the behaviour of the time derivative of the Wigner eigenfunctions for each $i$-sector of the $2$-dimensional harmonic oscillator.

\section{Discussion and Conclusions}
\label{sec:Conclusions}

Let us briefly discuss our results. First of all, it is natural to expect that for too small values of $\gamma /\Omega$ (say $\gamma /\Omega \ll 0.002$), time crystal and NC beating patterns are hard to measure. Nevertheless, it is 
interesting that 2-dimensional (or 3-dimensional) Bose-Einstein condensates with time crystal like behaviour have been detected through a resonance between two (or three) oscillating mirror atoms \cite{Sacha0,Sacha1}. Our results suggest that the observed behaviour is a natural explanation for the quasi-periodic eigenstates driven by $\xi_1$, $\xi_2$, and $\gamma$. In fact, for extended time intervals, in which the NC quantum beating takes place, the short time scale $\Omega$-frequency periodic behaviour turns into a quasi-periodic one, due to the periodic corrections from $\gamma$-frequency. 
This suggests that a connection of our results with the spontaneous formation of time quasi-crystals from atoms bouncing between a pair of orthogonal mirror atoms \cite{Sacha0,Sacha2} is possible.

On general grounds, our results show that the non-stationary behaviour associated to time crystals, arises entirely from either position ($q$) or momentum ($p$) noncommutativity, i.e., from $[\hat{q}_i,\,\hat{q}_j]\neq i \theta \epsilon_{ij}$ and/or $[\hat{p}_i,\,\hat{p}_j]\neq i \eta \epsilon_{ij}$, with no need of an {\it ab initio} hypothesis of spontaneous breaking of time translation symmetry.
Thus, we can conclude that the NC parameters naturally give origin to periodic oscillations that resemble time crystals. Conversely, besides accounting for the emergence of such unexpected properties, the measurable oscillation amplitude $\propto \hbar \gamma \Omega$ ($\gamma/\Omega$ driven by the NC parameters, $\gamma$, and the external oscillation frequency $\Omega \sim \omega$, can themselves be tested in order to set bounds to the NC parameters. Thus, we hope that our discussion in the context of the 2-dimensional noncommutative quantum harmonic oscillator might stimulate further attempts to experimentally test such a fascinating phenomena as time crystals. In fact, recent claims on the observation of continuous time crystal behaviour in quantum processors \cite{Mi,Frey}, atom-cavity system excited by photon oscillations \cite{Kongkhambut} and in an electron-nuclear spin 
system \cite{Greilich} show that there is a vivid interest in searchimg for concrete experimental evidence of time crystals in Nature.

As a final remark, we point out that some discrete or Floquet time crystals have been considered, for instance, in finite dimensional Hilbert spaces \cite{Flo1,Flo2}, however these might not be related to the ones discussed here, which emerge from the continuous deformed algebra, Eq. (\ref{EQEq31}).

\noindent
{\bf Acknowledgments} 

\noindent
The work of one of us (O.B.) is partially supported by FCT (Fundação para a Ciência e Tecnologia, Portugal) through the project CERN/FIS-PAR/0027/2021, with DOI identifier 10.54499/CERN/FI.

%%%%%%%%%%%%%%%%%%%%%%%%%%%%%%%%%%

\end{document}